\documentstyle[11pt]{article}
\begin{document}
{\bf Accepted to Physical Review B}
\begin{center}
{\Large The phase-separated states
 in antiferromagnetic semiconductors with polarizable lattice  }\\
{E.L. Nagaev}\\  
Institute for High Pressure Physics, Troitsk, Moscow Region,   
142092, Russia\\  
\end{center}  
  
\begin{abstract}
The possibility of the slab or stripe phase separation (alternating
  ferromagnetic  highly-
conductive  and insulating antiferromagnetic layers)  is 
proved for isotropic degenerate antiferromagnetic semiconductors.  
This type of phase separation competes with the droplet phase 
separation ( ferromagnetic droplets in the antiferromagnetic host or 
vice versa). The interaction of  electrons with optical phonons alone 
cannot cause phase-separated state  with alternating highly-conductive 
and insulating regions but it stabilizes the magnetic phase separation.
The magnetostriction deformation of the lattice in the phase-separated 
state is investigated.
\end{abstract}

\section{Introduction}

In magnetic semiconductors there are two main mechanisms   of the 
charge carrier self-trapping:  magnetic and polaronic. The former 
corresponds to trapping of the electron by a microregion of a changed 
magnetic phase created by this electron. This effect is most 
energetically favored in  antiferromagnetic semiconductors where the 
self-trapping occurs in  ferromagnetic microregions$^{1,2}$. On the 
other hand, the magnetic semiconductors are partly  polar crystals.
For this 
reason  polaronic effects should exist in them. The elecron-phonon 
interaction is not so strong in them as in the completely polar 
 NaCl-type crystals, and  only it alone cannot change 
the state of the charge carriers drastically. In particular,
the small polarons are impossible in the magnetic semiconductors$^3$.
Otherwise, they would not be degenerate at heavy dopings, and the charge
carriers would not realize the indirect exchange in them. 
 But,  being added to the 
magnetic mechanism of the self-trapping, the  lattice mechanism   can 
considerably increase the stability of the self-trapped state$^{4}$.  

In degenerate antiferromagnetic semiconductors the  magnetic self-
trapping is a  cooperative phenomenon. It manifests itself as a phase 
separation into the highly-conductive  ferromagnetic phase and 
insulating phase as a result of the fact that all the 
charge carriers concentrate inside the  ferromagnetic phase$^{5,6}$. Both 
these phases are charged oppositely.   Due to the  Coulomb interaction 
between them the phase intermixing takes place.
	
 Two main types of such intermixing were considered  in 
an isotropic crystal: the droplet one$^6$
 and the slab one$^5$. In the former case 
at relatively low charge carrier  densities, droplets of a metallic 
ferromagnetic phase arise inside an insulating antiferromagnetic host, 
with all the charge carriers concentrated inside these droplets. At 
larger densities, the highly-conductive  ferromagnetic phase occupies 
the larger part of the volume, and the antiferromagnetic phase is 
represented by  insulating droplets inside it. The slab structure 
corresponds to alternating highly-conductive  ferromagnetic and 
insulating  antiferromagnetic plane layers. As a particular case,
it includes the stripe structure with the ferromagnetic phase consisting
of monolayers which separate the antiferromagnetic slabs. As is well known,
the stripe structrure attracts much attention in connection with
the manganites, nickelates, and high $T_c$ superconductors nowadays.

In the pioneering paper [5]  
only very  crude estimates of the energy of the slab phase-separated state   
were carried out.  The electric properties of such a  phase-separated 
state should be highly anisotropic: it should be highly-conductive 
along the layers and insulating  across them for any
charge carrier  density in the range where the  phase separation  is 
realized.   In  the subsequent paper [6]    
 much more accurate calculations were carried for the droplet model. 
 This  
model made it possible to explain the transition from the insulating 
state to the highly-conductive state with increasing charge carrier  
density  as a result of the percolation of the ferro-electronic liquid, i.e. 
of changing the topology of the  ferromagnetic phase from multiply-
connected to  simply-connected. Such a transition  was observed   
experimentally in   EuSe and in other materials (see Refs. [3,7]).

The problem of the phase separation becomes again highly actual due 
to enhanced interest to the manganites. The coexistence of the  
ferromagnetic and antiferromagnetic phases in them was established 
very long ago$^8$. There are new experimental data  which evidence  
coexistence of these types of the magnetic ordering (e.g., Refs. [9-16]). 
Though  sometimes these data were interpreted in terms of  the canted 
antiferromagnetic ordering, one should keep in mind that it is unstable 
against fluctuations of the charge carrier  density$^{3, 17-21}$. For this 
reason one can assume that these properties are related to the ferro-
antiferromagnetic  phase separation. This  makes it necessary to 
reinvestigate theoretically  the  phase separation in the  degenerate 
antiferromagnetic semiconductors more accurately. In particular, it is 
necessary to find out whether the slab structure is really impossible 
in them.

 The present paper is devoted to this problem.  A more general model 
than in  Refs. [5,6] is used here which takes into account not only the 
magnetic self-trapping but also the polarization that. For this reason 
essentially new results are obtained not only for the slab structures 
but also for the droplet ones.	It will be shown that for some systems, 
both polarizable and nonpolarizable,  the difference in the energies of
 the slab and droplet structures is so 
small that it is below the accuracy of the calculation. Hence,  one 
cannot state with certainty that the slab structure is impossible in the 
isotropic antiferromagnetic crystals. 
The small difference in the energies assumes that  a small force can 
transform  one structure into the other even if the latter is normally 
unstable. For example, if normally the droplet structure is stable, a 
small uniaxial deformation can stabilize the slab structure.

Independently of the type of  the structure, the lattice polaronic effects  
can essentially influence the electron-magnetic  phase separation 
making it considerably more stable. They increase the size of the  
ferromagnetic regions in both droplet and slab structures. On the 
other hand,  they  increase the density of electrons in the   
ferromagnetic phase. As a result, the total volume of the  
ferromagnetic phase and the total magnetization of the crystal
 become decreased.  
Hence, the electron-lattice interaction, suppressing the  ferromagnetic 
phase,  hinders the  metal-insulator  transition   for the droplet 
geometry and the transition to the uniform ferromagnetic 
state for both geometries. 

Especially large  is contribution of the polaronic  effect in the case 
when the polaronic attraction between the charge carriers almost 
compensates the Coulomb interaction arising due to the mutual 
charging of both phases. If the former exceeds the latter, the phase 
separation of the types discussed here becomes impossible. The 
condition for this the inequality $\epsilon_0 /2 < \epsilon_{\infty}$. 
where $\epsilon_0$ and $\epsilon_{\infty}$ are  the static and high-
frequency dielectric constants, respectively.
But such a situation will not be considered here. There are 
antiferromagnetic semiconductors, in which $\epsilon _0/2$ only 
slightly exceeds $\epsilon_{\infty}$, e.g., EuSe with $\epsilon_0$ = 9.4 
and $\epsilon_{\infty}$ = 5.0  [22], so that the investigation of the 
polaronic effects close the stability boundary   is a quite actual 
problem.

Recently some authors expressed the point of view that the purely 
polaronic phase separation is possible in the manganites without 
magnetic phase separation (e. g., [23]). This seems hardly possible as 
such a state would correspond to formation of regions consistsing of 
the Pekar polarons. But, as is well known, the Pekar polarons are 
impossible for realistic parameter values. This qualitative 
consideration was confirmed numerically  using the procedure 
described below. The situation can be different when the small
polarons are more energetically favored than the large polarons.
Accordong to Ref. [24], the small polaron condensation into a string 
is possible. But, as was already mentioned, the small polarons do not
exist in the magnetic semiconductors.

	 It should be noted that other types of the electron-phonon 
interaction can be essential 
for certain systems. In Ref. [25]  the idea was advanced that the 
interaction between the electrons and 
Jahn-Teller phonons  can cause the phase separation 
in the manganites.  But  this 
problem will not discussed here.  In Ref. [26] the magnetostriction at  
the magnetic single-electron self-trapping  was investigated. In the 
present paper the modulation of the lattice due to the interaction of the 
slab structure with the acoustical phonons will be investigated. The 
problem of magnetostriction at the electron self-trapping became 
actual now due to its experimental discovery  [27, 28].

\section{The Hamiltonian of the system}
	
	The Hamiltonian of the system consists of the Hamiltonian of 
the $s-d-$ model $H_{sd}$, the free phonon Hamiltonian $H_{p}$ 
and the Hamiltonian $H_{sp}$ of the $s$-electron-phonon interaction.
$$H = H_{sd} + H_{sp} + H_{p} + H_C \qquad (1)$$
$$H_{sd} =  t \sum a^{*}_{{\bf  g}, \sigma }a_{{\bf g + \Delta}, 
\sigma}
-A \sum \left({\bf sS_{g}}\right)_{\sigma \sigma'}
 a^{*}_{{\bf g} \sigma} a_{{\bf g}\sigma'}$$

$$- \frac{I}{2} \sum {\bf S_{g}S_{g+\Delta}} \qquad (2)$$

where  $a^{*}_{{\bf g} \sigma},  a_{{\bf g+\Delta}\sigma}$
   are the $s$-electron operators, corresponding to the conduction 
electrons 
or  holes  at the atom {\bf g} with the spin projection $\sigma$, 
{\bf s} is the $s$-spin operator, ${\bf S_{g}}$  that of the $d$-spin 
of the atom {\bf g}, ${\bf \Delta}$ the vector connecting
 the first nearest neighbors.  The crystalline structure is assumed
 to be simple cubic, the $d$-spin magnitude being $S$. As usually, the 
inequality $t  \gg  IS^2$
must be met as the hopping integral $t$ is of the first order of 
magnitude, and  the $d-d$- exchange integral $I$ of the second order 
in the small $d$- orbital  overlapping.

The free phonon Hamiltonian for essential phonons is given by the 
standard expression
$$H_p = \sum \omega_{\bf q}b^{*}_{\bf q}b_{\bf  q} \qquad  (\hbar 
= 1) \qquad (3)$$
where $ b^{*}_{\bf q},  b_{\bf q}$ are the operators of the phonons 
with the wave vector ${\bf  q}$. 

The electron -  phonon  Hamiltonian can be presented in the form
$$H_{sp} = - \sum [C_{\bf q}{\rm exp}(i{\bf q g}) b_{\bf q} 
a^{*}_{{\bf g} \sigma} a_{{\bf g}\sigma} + {\rm H.C.}]  \qquad 
(4)$$
In the long-wave approximation,  the interaction constant with the 
longitudinal optical  phonons is given by the expression
$$C_{\bf q}^o = -i e \sqrt{\frac{2 \pi \omega }{\epsilon^{*} V 
}}\frac{1}{q}, \qquad \frac{1}{\epsilon^{*}} = 
\frac{1}{\epsilon_{\infty}} - 
\frac{1}{\epsilon_{0}}\qquad  (5a)$$
 where $\omega$ is 
the longitudinal optical phonon frequency, $V = N a^3$ is the volume 
of the crystal, $a$ is the lattice constant. 
For the longitudinal acoustical  phonons   in the deformation potential  
approximation  the electron-phonon constant is

$$C_{\bf q}^a =  iE_1 q \sqrt{\frac{ k^2}{2 \omega_{\bf q}M 
N}} \qquad \omega = s q \qquad(5b)$$
where $E_1$ is the deformation potential constant,  $M$ the mass of the 
unit cell.

The term $H_C$ describes the Coulomb energy of interaction of the $s$-
electrons with each other and with ionized  impurities. The charge of 
the latter compensating  the charge of the charge carriers is assumed to 
be distributed uniformly over the sample (the jellium model).

\section {The variational procedure}

	The reason for the magnetoelectronic phase separation in an 
antiferromagnetic semiconductor  is the fact that the charge carrier 
energy is lower for the  ferromagnetic ordering than for the 
antiferromagnetic that. If the charge carrier  density is insufficient to 
make the entire crystal  ferromagnetic, the charge carriers can 
concentrate in a portion of the crystal and make it  ferromagnetic.

. Obviously, the  ferromagnetic and antiferromagnetic phases are 
charged oppositely. To diminish the Coulomb energy, these phases 
should intermix. Two geometries will be considered. The first of them 
assumes that a crystal becomes separated into alternating layers of a 
highly-conductive  ferromagnetic phase and  the insulating 
antiferromagnetic phase$^5$. The second is the droplet geometry of the 
phase separated state investigated in Ref. [6]. Only the charge carrier  
densities below the percolation threshold will be  considered here 
which corresponds to the  ferromagnetic highly-conductive droplets 
inside the antiferromagnetic host. 
	
Like in Refs [5,6], a variational procedure  will be used here to find the 
ground-state energy of the phase-separated system. One of the 
variational parameters is the ratio of the volume of the antiferromagnetic 
phase to that of the  ferromagnetic phase $X$.  For the slab geometry 
the second variational parameter is the thickness of  the  ferromagnetic 
layers  $L$ with the antiferromagnetic layers of a thickness $XL$ (in 
the lattice constants $a$). Both numbers $L $ and $XL$ are assumed to 
be integer. Then, if the lattice is simple cubic, the electrons are 
concentrated in the  ferromagnetic layers with dimensionless 
coordinates $f_x = nL(1+X)+  g_x, ( 1 \le g_x  \le L)$ where their 
number per atom $\nu(x)$ is equal to $\nu_0 (1+X)$ ($\nu_0$ is the 
mean number of $s$-electrons per magnetic atom, $n$ an integer). Outside these 
layers the number of the conduction electrons is zero. For the droplet 
geometry  the second variational parameter is the radius of the  
ferromagnetic droplet $R$.

In the  ferromagnetic phase all the charge carriers are assumed to be  
spin-polarized. For  $W = 12 t \ll AS$ this condition is met for all the 
charge carrier  densities. For $W = 12 t \gg AS$ it is met when $AS > 
\mu$ where $\mu$ is their Fermi energy.  

Two approaches will be considered here. The first one,  sketched in 
Ref. [5] and  used in Ref [6], is valid for  large $L $ or $R$ . In this 
case, in addition to the standard bulk kinetic energy $E_B$, the 
surface electron energy $E_S$ must
be introduced  which  stems from spatial quantization of  the electron 
levels. Just competition of the surface energy and the Coulomb 
energy  ensures a finite size of the  ferromagnetic regions. The second 
approach assumes that   a $s$-electron is confined to a single atomic 
plane, i.e., that its motion is  two-dimensional ($L$ = 1).     
The  ground-state total energy of the phase-separated system 
$E_{PS}$ (per  unit cell) can be represented in the form:
$$E_{PS}  = E_k  -   \frac{AS\nu_0}{2}+ E_C  + E_{dd}+ E_{pol} 
\qquad (6)$$
For large $L$ or $R$, the kinetic energy of the system can be 
presented in the form
$$E_k = E_B +E_S \qquad (7)$$
Here the bulk kinetic energy of the charge carriers $E_B$ is given by 
the expression valid for relatively small $\nu$:
$$E_B =  - 6t \nu_0 + K (1+X)^{2/3}, \qquad K = 0.6 \mu(\nu_0) 
\nu_0, \qquad  \mu(\nu) = t(6\pi^2 \nu)^{2/3} \qquad (8)$$.
where $\nu_0$ is the mean s-electron number per unit cell.

The term $E_S$ is the surface energy of the electron gas. Like in Ref. 
[6], it is calculated in the Born-Oppenheimer approximation, i.e. by 
expansion in  $1/[\nu_0L(1+X)]$ powers. The Dirichlet boundary 
conditions for the $s$-electron wave function are taken. For the slab 
structure
$$E_S^s = \frac{2^{2/3}5 \pi^{1/3} K (1+X)^{1/3}}{ 
3^{1/3}16\nu_0^{1/3}L } \qquad (9a)$$    
and for the droplet structure
$$E_S^d =  \left (\frac{\pi}{6}\right)^{1/3}\frac{15 K}{16 
\nu_0^{1/3}(1+X)^{1/3}R} \qquad (9b)$$

For the case of $L $ = 1 one obtains the following expression for the 
total kinetic energy
$$E_k^1 = - 4t \nu_0 + 2 \pi \nu_0^2 (1+X)t \qquad (10)$$

The term $E_C$ in Eq (6) representing the Coulomb energy of the 
phase- and charge-separated state is calculated using the jellium 
model:  the total compensating charge with the density (- $e \nu_0$) 
modeling the charge of ionized donors or acceptors is distributed 
uniformly over the crystal.  For the slab structure one has:
$$E_C^s = \frac{\pi e^2 \nu_0^2 L^2 X^2 }{6 \epsilon_0 a} \qquad 
(11a)$$

For the droplet structure the calculation is carried using the Wigner 
spheres of  a radius $R_W$ surrounding the  ferromagnetic droplets. 
The total charge of the Wigner sphere is zero so that
$R_W = R(1+X)^{1/3}$. 
Hence,
$$E_C^d = \frac{2 \pi  e^2 \nu_0^2 R^2 [2X +3 - 3(1+X)^{2/3}]}{5 a 
\epsilon_0} \qquad (11b)$$

The  change in the $d-d$-exchange energy due to the formation of the  
ferromagnetic regions is given by 
$$E_{dd} = \frac {D}{1+X}, \qquad D = -z IS^2 \qquad (12)$$
where $z = 6$ is the coordination number for a simple cubic lattice, $I 
< 0$.

\section{The polaronic energy}

 The polaronic energy $E_{pol}$ entering Eq. (6) 
is found from
 the effective lattice Hamiltonian    
  $H_{pol} = H_p + <H_{sp}>$ where the symbol $<...>$ denotes averaging 
over the ground state of the electronic-magnetic subsystem. This 
corresponds to the adiabatic approximation in the polaron theory 
which is justified even for  a weak electron-phonon coupling since the 
nonuniform electron distribution arises not due to the phonons  in a 
selfconsistent manner  but due to the magnetization. 
This distribution is given in advance with respect to the
phonons.  

 For the slab structure on obtains from Eq. (4):
$$<H_{sp}>  = -\sum_{\bf  f, q} C_{\bf  q}\nu (f_x) [b_{\bf q} {\rm 
exp}(i {\bf q f}) + {\rm H.C.}], \qquad  (13) $$
$$\nu(f_x)  = \nu_0 (1+X) \sum_{n}  \Theta [f_x - n(1+X)L]  \Theta 
[n(1+X)L + L - f_x], \qquad (14)$$
where $\Theta (z) = 1$ for  $z > 0$ and  $\Theta (z) = 0$  for $z \le 0$.

One obtains from Eqs (13), (14):       
$$<H_{sp}>  = - i N_y N_z \nu_0 (1+X) \sum_{q_x} 
[C(q_x, 0, 0)  b_{q_x ,0,0}P(q_x)p(q_x) + {\rm H.C.}], \qquad (15)$$
 $$p(q_x) = \frac{{\rm exp}(i q_x L) - 1}{{\rm exp}(i q_x ) - 1}, 
\qquad (16)$$
$$P(q_x) =  \frac{{\rm exp}[i q_x G L (1+x)] - 1}{{\rm exp}[i q_x L 
(1+x )] - 1}, \qquad (17)$$
where
$$G = \frac{N_x}{L(1+X)} \qquad (18)$$
 is the number of pairs of the  ferromagnetic and antiferromagnetic 
layers, $L(1+X)$ the dimensionless thickness of  a pair of these layers. 
As $G \to \infty$,  the quantity $P(q_x)$ is nonzero only for  
$q_{m} = 2 \pi m/[a (1+X)L]$  where $m$ is an integer. Hence,
$$P(q_x)  = G\sum_{m= - Y}^{Z} \delta\left(q_{x}a, \frac{2 \pi 
m}{(1+X)L}  \right), \qquad (19)$$
$$ \qquad  Y = \frac{L(1+M)}{2} + \frac{1}{4}[ {\rm cos}(\pi(1+X)L) - 1];   
 \qquad Z =  Y - 1\qquad(19)$$
where $\delta(n,m) = 1$ for $n = m$ and 0 otherwise.

Using Eqs (16), (17) and diagonalizing the averaged phonon 
Hamiltonian $H_{pol}$ (3), (13 - 15)  by the well known procedure of the 
phonon operator shift  
 $$b_{\bf m} \to  b_{\bf m} + \beta_{\bf m}, \qquad \beta_{\bf  m} =
 \frac{N \nu_0 C_{\bf m}p(q_{m})}{L \omega_{\bf m}}, \qquad 
(20)$$
$${\bf m} = \left(\frac{2 \pi m}{a(1+X) L}, 0, 0\right), \qquad N = 
N_xN_yN_z$$
one obtains the following expression for the cooperative polaron 
energy per unit cell:      
$$E_{pol}^s = - \frac{N \nu_0^2}{L^2} \sum_{m = -Y}^{Z} \frac{|C_{\bf 
m}|^2 |p(q_{m})|^2}{\omega_{\bf m}}, \qquad (21)$$  
$$|p(q_{ m})|^2  = \frac{1 - {\rm cos}2\pi m/(1+X)}{1 - {\rm cos}2\pi 
m/[(1+X)L]} $$
The term with $m = 0$ is omitted from Eq (21) as it is compensated 
by interaction of phonons with the ionized impurity charge.

For the droplets  such an accurate calculation of $E_{pol}$ will 
not be carried out for a reason pointed out in the end of this Section.  
It is sufficient to  evaluate this quantity again using 
the approach based on the Wigner spheres. 
 Then in the
 adiabatic approximation (e.g., Ref. [3])
$$E_p^d = - \frac{1}{8 \pi \epsilon^*}\int D^2({\bf r}) d^3r \qquad  
(22)  $$
where $D({\bf  r})$ is the electric induction.
Calculating this quantity for each Wigner sphere and multiplying it by 
their number, one comes to the expression for $E_{pol}^d$.

 To determine $E_p^s$ using the same simplified treatment, one 
should note that one can introduce an analogue of the Wigner  spheres 
for the slab structure, too. Namely, one can introduce the Wigner 
layers with the total zero charge, consisting of the highly-conductive 
layer  with the $x$ coordinates between $-L/2$ and $L/2$ and two adjacent 
insulating half-layers between $L(1+X)/2$ and $L/2$ and symmetrical 
one.  For them in the adiabatic approximation (22) one arrives to an 
expression  for  $E_{pol}^s$  similar to that for $E_{pol}^d$. For them both 
$$E_p = - E_C  \frac{\epsilon}{\epsilon^*} \qquad (23)$$                                    
The question arises about the accuracy of Eq (23). One observes that  
for the slabs one obtains the same expression directly from Eqs (21), 
(11a) if the quantities $L$  and $X$ tend to $\infty$ and      
to 0, respectively. The same occurs for any $L$  as $X  \to \infty$. 
Really, in this case the sum over $m$ in Eq (21) becomes equal to $2 
X^2 L^2 \zeta(2)$ where $\zeta(n)$ is the Riemann zeta-function. 
Hence, Eq (23) and its analogue are valid for  large-scale phase 
separation with relatively small  increase in the density or  for very 
large volumes of the antiferromagnetic phase independenty
of the size of the ferromagnetic layers. As follows from Eqs 
(11a,b) and (23), the polaron instability ($E_C$ + $E_p < 0$) should 
occur  for
$$\frac {\epsilon_0}{2} < \epsilon_{\infty} \qquad (24)$$

If these conditions are not met,  Eqs. (21) and  (23)  lead to different 
results. For example, if one takes $\epsilon_0$  = 5 and  
$\epsilon_{\infty}$ = 2.5,  then  one obtains by numerical calculations  
from Eqs (21) and (11.a):  $E_p/E_C$ = 1.0008  for $L$ = 1 and     $k 
\equiv  XL = 1000$ or  for $L$ = 1000 and $k$ = 1 in  agreement with Eq 
(23).  But for $L = k = 1$ one obtains a value of 2.4317  for this ratio. 
This means that for such a structure the polaronic effects are considerably 
stronger than for  thicker  ferromagnetic or antiferromagnetic 
layers. Hence, in principle, they can stabilize the stripe
  structure when the ferromagnetic layers are reduced to
monoatomic planes  (the long-rang ferromagnetic ordering is possible
in the two-dimensional systems even at $T \ne 0$ due to the magnetic
 anisotropy).   

As is seen  from Eqs   (23) and (11a,b),  in the adiabatic 
approximation used here the polaronic energy vanishes for the uniform  
ferromagnetic state  (for the slab structure it follows from the fact 
that  according to Eq (21) $p(q_m) = 0$ for $X = 0$).  

Now we return to the problem of a calculation of the polaronic energy
in the manner similar to that leading to Eq. (21). As  follows from
above-said, such a calculation could be very useful for  monoatomic
sublattices which are the limiting case for the droplet model.
But such sublattices have nothing in common with the phase separation
as single atoms cannot form a ferromagnetic phase. Thus, this problem
is beyond the framework of the present paper, and for this reason
is not considered here. The Eq. (23) is sufficient for the droplet
model which assumes that the size of the ferromagnetic droplets is
 large compared with the lattice constant.

\section {Calculations and their results}
 
	Now  results of calculations will be presented. First, it should 
be noted that a necessary condition for a  phase-separated state to be 
energetically favored consists in the requirement that the energy (6) 
be less than the energy  of the uniform  antiferromagnetic state 
$E_{AF}$:
$$E_{AF} = -6t\nu_0 + 2^{-2/3}K \qquad {\rm for} \quad W \gg AS 
\qquad  (25)$$
$$E_{AF}^{de} =  - \frac{6t\nu_0}{\sqrt{2S+1}} + 
\frac{K}{\sqrt{2S+1}} - \frac{AS \nu_0}{2}  \qquad {\rm for} \quad 
W \ll AS \qquad (26)$$
The latter Equation  for the double-exchange system is obtained using 
an expression for the charge carrier energy in a system  with the 
double exchange, if the $d$-spins are quantum [3]. It should be noted 
that the term $AS/2$ does not disappear  for  any type of the magnetic 
ordering playing the part of  an additive constant in the case of the 
double exchange. 

According to Eq (23), if one wishes to account for the polaronic 
effects,   in the case of sufficiently large radii $R$ or layer thicknesses 
$L$   one can introduce the  effective dielectric 
constant  $\kappa$, defined by the relationship
$$\frac{1}{\kappa} = \frac{1}{\epsilon_0}- \frac{1}{\epsilon^{*}} =
\frac{1}{\epsilon_0} -  \frac{2}{\epsilon_{\infty}} \qquad (27)$$
In this case one can carry out the minimization of the energies (6-9a,b
),(11 a,b), (12)  
with respect to $R$ or $L$ in an explicit form, and then only the 
minimization with respect to $X$ should be carried numerically. If one 
interested only in the pure magnetic self-trapping, one should leave Eqs 
(11a,b) as they stand, and if one wishes to take into account the polaronic 
contribution, one should replace $\epsilon_0$ by $\kappa$ (27) in them.

In this case one can exclude the quantities $R$ ot $L$  from the 
expressions (6)   for the energy and make it a function of the only 
variational parameter $X$.  For the droplets structure with both  
ferromagnetic and antiferromagnetic droplets this procedure was 
carried out in Ref. [6]. For the slab structure this program was not 
carried out yet.

For the droplet model, as  in Eq (6) only the terms $ E_S^d$ and  $E_C^d$
 are $R$-
dependent,   one should replace the sum $E_S^d + E_C^d$ by its 
minimal  value with respect to $R$:
$$E_ {SC}^d  \cong 1.2 \nu_0 \gamma [2X + 3 - 
3(1+X)^{2/3}]^{1/3}(1+X)^{2/9} \qquad (28)$$
$$\gamma =\left [\frac{\mu(\nu_0)^2 e^2 \nu_0^{1/3}}{\epsilon_0 
a}\right]^{1/3}$$
and substitute the following expression for  the radius:
$$R = \left\{\frac{75 K (1+X)^{1/3} \epsilon_0 a}{6^{1/3} 64 
\pi^{2/3} \nu_0^2e^2 [2X + 3 - 3(1+X)^{2/3}]} \right\}^{1/3} \qquad 
(29)$$
A similar calculation for the slab structure leads to the expressions
$$E_{SC}^s = \frac{3}{2}\left(\frac{\pi}{3}\right)^{1/3}\left(\frac{ 
2^{2/3}\pi^{1/3}}{16}\right)^{2/3}K^{2/3}\left(\frac{e^2}{\epsilon_
0 a}\right)^{1/3} \nu_0^{4/9}x^{2/3}(1+x)^{2/9} \qquad (30)$$
$$L = \left[ \frac{ 2^{2/3}15 (1+X)^{1/3} \epsilon_0 a}{16 
\pi^{2/3}e^2 \nu_0^2 X^2}\right]^{1/3} \qquad (31)$$
Further  calculations should be carried numerically.

Now some results of them will be presented. First, we consider the 
case when the the system is close to the boundary of  the polaron instability. 
For this aim the same  values of the dielectric constants will be taken 
as in EuSe [23]: $\epsilon_0 = 9.4$, $\epsilon_{\infty}= 5$. As EuSe is 
an isotropic metamagnet with a very low critical field of about 2000 
Oe,  a value of $D$ = 0.0001 is chosen (here and below all the typical 
energies are in the eV units). Other quantities are:  $t$ = 0.2, $A$ = 
1.2/7, $e^2/a$ = 2.7,  $S$ = 7/2 . Not all they coincide with the 
corresponding quantities in EuSe though are likely to be  close to 
them.  Nevertheless, the inequality $W \gg AS$ is certainly met here, 
and the energy of the system will be counted off from the expression (25).

	The results of the calculations are depicted in Fig.1 for both
 droplet and slab structures. Here and below the upper indices
 $p$ and $r$ will denote
the polarizable and rigid lattices, the lower indices
$d$ and $s$  the droplet and slab structures, respectively.
 The parameters   $X_s^p$, $X_d^p$,  $R^p$ and $L^p$ are 
presented  in Fig.1 
 For the droplet model the calculations are carried only
down to the $X^p$ value of 1, which corresponds to the electronic and 
 ferromagnetic percolation.
Calculations of the polaronic effects for $X^p < 1$ should be carried out using
another procedure.

 As is seen from  Fig. 1,  the parameters $X^p_d$ and $X^p_s$  decrease
 and hence  the ferromagnetic portion of the crystal increases with increasing
charge carrier  density $\nu_0$.  Respectively,
 the sizes of the  ferromagnetic 
regions $R^p$ and $L^p$ increase. For the slab structure
 an abrupt concentration
phase transition into the uniform  ferromagnetic state takes place though the
$X^p$  jump is rather small. As the polaronic contribution to the energy
 is small
at $X^p  \to 0$,  this effect is not related to the  polaronic effect.
 A corresponding 
calculation for the unpolarized  droplet model using results of
  Refs [3,6] shows that the
 phase separation  also disappears abruptly. 

Comparing numerical results for these two  models, one should point out
that their energies and $X^p$  values are close to each other. In both cases the 
energy per  conduction electron decreases smoothly with increasing electron
 density.  For the droplet structure this energy $E^p_d$
 changes from - 0.2089
 at $\nu_0$ = 
4 $\times 10^{-4}$  ($X^p_d$ = 5) to - 0.2245  at $\nu_0$ = 
1.2 $\times 10^{-3}$  ($X^p_d$ =  1). For the slab structure this energy
$E^p_s$ is
 equal to - 0.2058
 at $\nu_0$ = 4$\times 10^{-4}$  ($X^p_s$ = 4.6), - 0.2256 at $\nu_0$ = 
1.2 $\times 10^{-3}$  ($X^p_s$ = 1),  and - 0.2387  at $\nu_0$ = 
1.98 $\times 10^{-3}$  ($X^p_s$ =  0).  It is interesting to note that
 the densities 
corresponding to equal volumes of the antiferromagnetic and  ferromagnetic 
phases coincide for both models.

One should compare results obtained with accounting and 
without accounting  for the lattice polarization. It is advisable
to carry out this comparison for the density   
$\nu_0$ = 0.00095 corresponding to the percolation threshold
 in the rigid lattice.
One obtains: for the droplet model in the rigid lattice 
the energy per electron  $E_d^r$ = - 0.2054, 
$X^r_d$ = 1.0, $R^r_d$ = 10, the number of electrons in the droplet 
$N_d^r$ = 8,3. The small parameter of theory ensuring applicability of Eq. 
(9a)  $s^r = 1/[(6\pi^2)^{1/3}\nu_0^{1/3} (1+x)^{1/3}]$ is 0.21 which 
justifies use of  the many-electron approach adopted here.
 With accounting for the the polaronic effects one has: the energy
per electron $E^p_d$ = -
0.2202, $x^p_d$ = 1.6, $R^p$ = 16, $N_d^p$ = 44, $s^p$ = 0.12.

Obviously, the polarization increases the number of the electrons in 
each droplet by a factor of  4.75, whereas its volume is increased only   
by a factor of 4.1.
 Hence, the polarization increases the density of 
conduction electrons. 
 The total magnetization of the crystal is proportional  to the relative 
volume of the  ferromagnetic phase $1/1+X$. Thus, the polarization 
decreases the total magnetization by about 30 \%. Simultaneously, the 
system is shifted from the percolation threshold to the insulating side.

Now these results will be compared with those obtained for the slab 
model with the same $\nu_0$ value of 0.00095.  Without polarization, 
one obtains: $E_s^r$ = -0.2073, $X_s^r$= 1.0, the  ferromagnetic layer 
thickness $L^r$ = 8.5. With polarization, $E^p_s$ = -0.2208, $X^p_s$ 
= 1.5, $L^p$ = 13. Though, formally, the energy of the slab 
structure is less than that of the droplet structure, the difference 
between then amounts only to 0.3 \%  which is certainly considerably 
less than the accuracy of the present calculations. For this reason it is 
impossible to decide which structure should be realized in reality.
 The reduction of  the magnetization caused by the lattice  polarization 
 in the slab structure is close to that in the droplet structure. 

Now we consider a double-exchange system of the manganite type 
with values of the dielectric constants coinciding with those of the La-
based manganite: $\epsilon_0 =5, \epsilon_{\infty} = 3.4$ [29]. The $s-d$-
exchange integral $A$ is assumed to be very large (its exact value is  
inessential as its enters the expression for 
the energy only as an additive constant ), and the following values will be taken: $t = 0.2, D = 0.01, S = 2$.
Then  one finds for the density $\nu_0 = 0.015$ close to the 
percolation density: for the  unpolarized droplet structure
 the energy $E^r_d$ = - 
0.06 is minimal for $X^r_d$ = 1.1  with $R^r$ = 4.2, $N_d^r$ = 10 and $s^r$ 
=0.19 (the energy is counted off from the energy $U$  (26)). If one takes 
into account the polarization, then $E^p_d$ = -0.09,  $R^p_d$ = 4.8,
 $N_d^p$ 
= 16, $s^p$  = 0.17 with $X_d^p$ = 1.3. 

In the case of the slab structure for 
the same values of parameters one finds: for a rigid  lattice $E^r_s$ 
= - 0.0708 with $X^r_s$ = 1.1 and $L^r$ = 3.44, for a polarizable lattice 
$E^p_s$ = -0.0986 with $X^p_s$ = 1.3 and $L^p$ = 3.84.

As is seen from these data, the main regularities established above for 
the case of the wide bands $W \gg AS$ remain in force for the case of 
the double exchange. The  difference in the energies of the slab 
and droplet structures is larger in this double-exchange system
 case which points to the droplet structure 
being preferential. One can also give examples where the situation is 
opposite. Being unable to predict in some cases which structure 
should be realized, I would like to point out that just due to this small
energy difference a small external force can cause transition of one structure
to the other. For example, if the droplet structure is normally realized
then a small uniaxial stress increasing the hopping integral $t$
 in the corresponding direction  can stabilize the slab structure.  

It is still necessary to discuss the possibility 
of the  ferromagnetic stripe structure ($L$ = 1). 
It is  clear that the single layer  ferromagnetic phase could be possible 
only in the  double exchange systems since in the opposite case $W 
\gg AS$ the energy of  the two-dimensional motion of the electron is  
by $2t = W/6$  higher than in the three-dimensional case. Meanwhile, 
the gain in  its energy due to the  ferromagnetic ordering is only  
$AS/2$. On the other hand, as was shown in the preceding section,  
the polaronic effect is enhanced in the structures with alternating ferro- 
and antiferromagnetic atomic planes. For this reason one can expect 
the existence of such structure in the double-exchange systems.

 I managed to prove the possibility of such structure in the limit $S \to 
\infty$. The following values of the parameters were used: $\epsilon_0 
= 5, \epsilon_{\infty} = 3.4, t = 0.1, D = 0.01$. Then one finds from 
Eqs (6) and (10)  that for  $\nu_0 = 0.02$ the system with $L^p$ = 1 is the 
most energetically favored for $k = 3$ (3 antiferromagnetic layers 
per  ferromagnetic layer). For $\nu_0 = 0.05$ the most energetically 
favored structure is  $L^p = 1, k =1$ with alternating ferro-and 
antiferromagnetic atomic planes. For small $S$, I did not 
find parameter values at which the stripe structure is most energeticlly
favored.
i

It remains only to consider the magnetostiction  deformation of a 
sample in the case of the slab  phase separated state.  For this aim 
one should use Eqs (20), (5b) yielding the shifts $\beta_{\bf q}$ of the 
operators for  the longitudinal acoustical phonons. Going  over from 
the phonon operators to the normal coordinates in the standard 
manner,one obtains the following expression for their shift $\delta 
Q_{\bf q}$: 
$$\delta Q_{\bf q} = \sqrt{\frac{1}{2 \omega_{\bf q}M}}[\beta_{\bf 
q} + \beta^{*}_{-\bf q}] \qquad (32)$$
where $M$ is the mass of the unit cell. Using this Eq, one obtains for 
the displacement of the atom {\bf g} along the $x$ axis related to the 
wave with the vector  ${\bf m}$ (20):
$$u_{\bf g} = A_{\bf m}{\rm exp}(i {\bf m g}), \qquad  A_{\bf m} = 
\frac{1}{\sqrt{N}} \delta Q_{\bf m} $$
 $$\frac{A_{\bf m}}{a} = \frac{E_1 m  |p_{\bf m}- p_{\bf m}^{*}| 
\nu_0 }{2M \omega^2 L} \qquad (33)$$
As follows from Eq (33), the lattice becomes modulated not only with 
the periodicity of the phase-separated structure but also the higher 
harmonics are present.

\section {Acknowledgments}

This investigation was supported in part by Grant No. 98-02-16148 of 
the Russian Foundation for Basic Research, NATO Grant HTECH LG 
972942  and  Grant INTAS-97-open-30253.

\section{References}
\
1. E.L.  Nagaev.   Pisma ZhETF  6,  484 (1967).

2.  E.L.  Nagaev.    Zh. Exp.Teor. Fiz.  54, 228 (1968).

3. E.L. Nagaev.  $Physics \quad of \quad Magnetic
\quad Semiconductors$. Moscow.  
Mir  1983

4. V.D. Lakhno, E.L.  Nagaev.  Fiz. Tverd. Tela. 18, 3429 (1976)

5. E.L.  Nagaev. Pisma ZhETF  16,  558 (1972).

6. V.A. Kashin, E.L.  Nagaev.  Zh. Exp.Teor. Fiz.    66,   2105 (1974)

7. E.L. Nagaev.  Uspekhi Fiz. Nauk.  165, 529 (1995);
Phys. Stat. Sol. (b) 186, 9 (1994)

8. E. Wollan, W. Koehler. Phys. Rev. 100, 545 (1955)

9. T. Okuda, A. Asamitsu, Y. Tomioka et al. Phys. Rev. Lett. 82, 3203 (1998)

10. T. Perring, G. Aeppli,  Y. Moritomo, Y. Tokura. Phys. Rev. Lett. 78, 3197 (1997)

11. R. Kajimoto, T. Kakeshita, Y. Oohara et al. Phys. Rev. B, R11837 (1998)

12. G. Papavassiliou, M. Fardis, M. Belesi et al. Phys. Rev. B 59, 6390 (1999)

13. H. Kawano, R. Kajimoto, M. Kuboto, H. Yoshizawa. Phys. Rev. B 53, 2202 (1996)

14. A. Vasiliu-Doloc, J. Lynn, A. Modden  et al. Phys. Rev. B 58, 14913 (1998)

15. D. Argyriou, J. Mitchell, C. Potter et al. Phys. Rev. B 56, 3826 (1996)

16. M. Hennion, F. Moussa, J. Rodrigues-Carvajal, L. Pinsard, and A. Revcolevschi, Phys. 
Rev. B 56, R497 (1997)

17. V.E. Zilbervarg and E. L. Nagaev. Fiz. Tverd. Tela 16,2834 (1974) 

18. E.L. Nagaev. Phys. Rev. B 58, 2415 (1998)

19. D. Arovas, F. Guinea. Phys. Rev. B 58, 9150 (1998)

20. M. Yamanaka, W. Koshiba, S. Maekawa. Phys. Rev. Lett. 81, 5604 (1998)

21. S. Mishra, S. Satpathy, T. Aryasetiawan et al. Phys. Rev. B 55, 2725 (1995)

22. M. Ikezawa, T. Suzuki. J. Phys. Soc. Jpn 35, 1556 (1973)

23. J. Zhou and  J. Goodenough. Phys. Rev. Lett. 80, 2665 (1998)

24. F. V. Kusmartsev. Phys. Rev. Lett. 84, 530 (2000)

25. S. Yunoki, A. Moreo and E. Dagotto. Phys. Rev. Lett. 81, 5612 (1998)

26.  V.D. Lakhno, E.L.  Nagaev. Fiz. Tverd. Tela. 20, 82 (1978)

27. J.M. De Teresa, M.R. Ibarra, P.A. Algarabel et al . Nature 386, 256 (1997)

28. R.V. Dyomin, L.I. Koroleva and A.M. Balbashov. Phys. Lett. a 231, 279 (1997)

29. K. Ahn and A. Millis. Phys. Rev. B 58, 3697 (1998)
$$  $$
{\bf  Captions}

Fig.1.  The ratio $X^p$ of the volumes of the antiferromagnetic
 and  ferromagnetic phases for
the droplet ($X_d^p$) and slab ($X_s^p$) structures
in the polarizable lattice, 
as well as the radius $R^p$ of the  ferromagnetic droplet for the former
 and the thickness $L^p$ of the  ferromagnetic layer
for the latter,  as  functions of the relative average charge carrier  density
  $\nu_0$.  $R^p$ and $L^p$  are expressed in the lattice constants.
 Numerical values of the parameters used, which correspond to $W \gg AS$, 
 are presented in the text just below Eq. (31).

\end{document}